\documentclass[12pt,preprint]{aastex}
\def\etal{{\it et~al.\,}}
\def\mic{$\mu$m$\,$}

\shortauthors{Hubeny, Burrows, \& Sudarsky}
\shorttitle{Strongly Irradiated Stars and Planets}

\begin{document}

\title{
A Possible Bifurcation in Atmospheres of Strongly Irradiated Stars
and Planets
}

\author{Ivan Hubeny\altaffilmark{1,2}}
\author{Adam Burrows\altaffilmark{3} \& David Sudarsky\altaffilmark{3}}

\altaffiltext{1}{NOAO, P.O.Box 26732, 950 N. Cherry, Tucson, AZ 85726}
\altaffiltext{2}{LASP, NASA Goddard Space Flight Center, Greenbelt, MD 20771}
\altaffiltext{3}{Department of Astronomy and Steward Observatory, The University
of Arizona, Tucson, AZ 85721}

\begin{abstract}
We show that under certain circumstances the differences
between the absorption mean and Planck mean opacities can lead to
multiple solutions for an LTE atmospheric structure. Since the absorption 
and Planck mean opacities are not expected to differ significantly
in the usual case of radiative equilibrium, non-irradiated atmospheres,
the most interesting situations where the effect may play a role are
strongly irradiated stars and planets, and also possibly structures
where there is a significant deposition of mechanical energy, such as
stellar chromospheres and accretion disks. We have presented
an illustrative example of a strongly irradiated giant planet where
the bifurcation effect is predicted to occur for a certain range
of distances from the star.
\end{abstract}


\section{Introduction}
There are many situations where the atmosphere of an object, 
a star or a planet, is irradiated by a companion star in such a way
that this irradiation significantly influences its atmospheric
structure. In the case of classical close binary stars, the effect exists,
but is rarely dramatic because the effective temperatures differ by
a factor of a few. On the other hand, the effect may be quite dramatic
in the case of sub-stellar mass objects such as giant planets and
brown dwarfs, irradiated by a solar-type star, in which case the
ratio of their effective temperatures may reach a factor of 100 or more.
We stress that we use the term {\it effective temperature} as
used in the theory of stellar atmospheres, namely
as a measure of the total energy flux coming from the interior of the
object.

In the case of extrasolar giant planets (EGP), we recently have witnessed 
a significant increase in interest in predicting EGP spectra. This was
motivated in part by two detected planetary transits, which are
in principle able to provide direct information about the planet's 
atmosphere. Another motivation is the need to predict
spectra of extrasolar planets to guide the design of future missions
which aim at recording EGP spectra .

We have recently computed a large set of model atmospheres and
spectra of EGPs (Sudarsky, Burrows, \& Hubeny 2003 - hereafter
referred to as SBH). We have used a modification of our universal stellar
atmosphere program TLUSTY (Hubeny 1988; Hubeny \& Lanz 1995),
called COOLTLUSTY. This variant
does not compute opacities on the fly; instead it uses a pretabulated
opacity as a function of wavelength, temperature, and density. The tables
are computed using an updated version of the chemical equilibrium 
code of Burrows \& Sharp (1999),
which includes a prescription to account for the rainout of species
in a gravitational field.

Motivated by the recent discovery of the second transiting planet
OGLE-TR-56 (Konacki et al. 2003; Sasselov 2003), 
which is believed to have a separation of a mere
0.0225 AU from its parent star, we tried to extend the SBH models
to higher irradiations. However, COOLTLUSTY faced significant convergence
problems. After various attempts and using various strategies, we have 
discovered that the convergence behavior is not the
result of a bug in the program, or of an insufficiency in our numerical
scheme, but is in fact a consequence of an existence of multiple solutions
that may lead to disastrous effects for convergence.
When studying the effect, we found that its roots are quite
general, and are in fact applicable to other cases in the 
theory of stellar atmospheres.
Therefore, we devote the present paper to explain the effect in general terms.


\section{Basic Equations}

The atmospheric structure is obtained by solving simultaneously
the radiative transfer equation, the radiative equilibrium equation,
and the hydrostatic equilibrium equation. For simplicity, we assume LTE.
Since our basic aim here is to study atmospheres of irradiated giant planets,
this approximation is a reasonable one, although it should be relaxed
in the future,  as was the LTE assumption eventually relaxed in modern
studies of stellar atmospheres. Here, we present a brief overview of
the basic equations.

The radiative transfer equation is written as
\begin{equation}
\label{rte1}
\mu \, \frac{dI_{\nu\mu}}{dm}=\chi_\nu\left(I_{\nu\mu}-S_\nu\right)\, ,
\end{equation}
where $I_{\nu\mu}$ is the specific intensity of radiation as a function
of frequency, $\nu$, angle - described through the cosine of the angle of propagation with respect to
the normal to the surface, $\mu$, and the geometrical coordinate,
taken here as the column mass $m$. The later is defined as
$dm=-\rho dz$, where $z$ is a geometrical distance (measured outwards),
and $\rho$ the mass density. The monochromatic optical depth is defined
as $d\tau_\nu = \chi_\nu\, dm$.
Finally, $S_\nu$ is the source function, given in LTE by
\begin{equation}
\label{sdef}
S_\nu=\frac{\kappa_\nu}{\chi_\nu}B_\nu+\frac{\sigma_\nu}{\chi_\nu}J_\nu\, .
\end{equation}
Here, $\kappa_\nu$ is the true absorption coefficient, $\sigma_\nu$ the
scattering coefficient, and
$\chi_\nu = \kappa_\nu + \sigma_\nu$ is the total extinction coefficient.
All coefficients are per unit mass.

The boundary conditions are provided through the diffusion approximation
at the deepest point,
\begin{equation}
I_{\nu\mu}(\tau_{\rm max}) = B_\nu(T(\tau_{\rm max})) +
\mu\, \frac{dB_\nu}{d\tau_\nu}|_{\tau_{\rm max}}\, ,\quad \mu > 0\, ,
\end{equation}
and the upper boundary condition is
\begin{equation}
I_{\nu\mu}(0) = I_{\nu\mu}^{\rm ext}\, , \quad  \mu< 0\, ,
\end{equation}
where  $I_{\nu\mu}^{\rm ext}$ is the specific intensity of the external
irradiation. For simplicity, we assume isotropic irradiation,
$I_{\nu\mu}^{\rm ext} = J_{\nu}^{\rm ext}$. It is convenient to express
the frequency-integrated irradiation intensity as
\begin{equation}
J^{\rm ext} \equiv \int_0^\infty J_{\nu}^{\rm ext}\, d\nu = W B(T_\ast)\, ,
\end{equation}
where $T_\ast$ is an effective temperature of the irradiating star
(in case the irradiation source is not a star, $T_\ast$ is merely the
characteristic temperature of the incoming radiation), and
$W$ a dilution factor.

The moments of the specific intensity are defined as
\begin{equation}
\label{momdef}
[J_\nu, H_\nu, K_\nu] \equiv (1/2) \int_{-1}^1 I_{\nu\mu}\, [1, \mu, \mu^2]\, d\mu
\, .
\end{equation}
The first moment of the transfer equation is written 
\begin{equation}
\label{Hmom1}
\frac{dH_\nu}{dm}=\chi_\nu\left(J_\nu-S_\nu\right)
=\kappa_\nu \left(J_\nu - B_\nu\right)\, ,
\end{equation}
where the second equality follows from equation (\ref{sdef}).
Integrating over frequency we obtain
\begin{equation}
\label{Hmomint}
\frac{dH}{dm} = \kappa_J J - \kappa_B B\, ,
\end{equation}
where $\kappa_J$ and $\kappa_B$ are the absorption and Planck mean
opacities, respectively, defined by
\begin{equation}
\label{kjdef}
\kappa_J = \frac{\int_0^\infty \kappa_\nu J_\nu d\nu }{\int_0^\infty J_\nu d\nu}\, ,
\end{equation}
and
\begin{equation}
\label{kbdef}
\kappa_B = \frac{\int_0^\infty \kappa_\nu B_\nu d\nu}{\int_0^\infty B_\nu d\nu}\, .
\end{equation}
These two opacities are usually assumed to be equal. This is an
excellent approximation in the case of non-irradiated atmospheres,
because $J_\nu \approx B_\nu$ in optically thick regions, and 
$J_\nu(\tau < 1) \propto (1/2) B_\nu(\tau=1)$ (the Eddington-Barbier
relation), so $J$ is proportional to $B$ and the averaging over
$J$ and $B$ leads to very similar results. 
However, here we maintain the
distinction because the difference between $\kappa_J$ and $\kappa_B$
turns out to be crucial in the case of strongly irradiated atmospheres.

The second moment of the transfer equation is 
\begin{equation}
\label{Kmom1}
\frac{dK_\nu}{dm}=\chi_\nu H_\nu\, ,
\end{equation}
and integrating over frequency we obtain
\begin{equation}
\label{Kmomint}
\frac{dK}{dm}=\chi_H H\, ,
\end{equation}
where
\begin{equation}
\chi_H = \frac{\int_0^\infty \chi_\nu H_\nu d\nu}
{ \int_0^\infty H_\nu d\nu}\, ,
\end{equation}
which is called the flux mean opacity. Notice that unlike the two
previous opacities, which were averages of the {\em true} absorption coefficient
(without the scattering term), 
the flux-mean opacity contains the {\em total} absorption coefficient.

Finally, the radiative equilibrium equation is written
\begin{equation}
\label{re1}
\int_0^\infty \kappa_\nu \left(J_\nu - B_\nu\right) d\nu = 0\, ,
\end{equation}
which can be rewritten, using the above defined mean opacities, as
\begin{equation}
\label{reint}
\kappa_J J - \kappa_B B = 0\, .
\end{equation}
Substituting (\ref{reint}) into (\ref{Hmomint}), we obtain another
form of the radiative equilibrium equation:
\begin{equation}
\frac{dH}{dm} = 0\, ,\quad {\rm or} \quad H={\rm const}\equiv (\sigma/4\pi)\,
T_{\rm eff}^4\, ,
\end{equation}
where $\sigma$ is the Stefan-Boltzmann constant.


\section{Temperature Structure}

\subsection{General}
\label{temp_gen}

The above equations are {\em exact}, given LTE, but are of course 
only formal
because the mean opacities $\kappa_J$ and $\chi_H$ are not known
in advance; only $\kappa_B$ is a known function of temperature and density. 
Nevertheless, assuming that they are known and equal to the Rosseland
mean opacity, we may write a solution
for temperature, following the classical textbook procedure, 
known by the name of ``LTE-grey model atmospheres'' (e.g. Mihalas 1978). 
A generalization of a classical
model for the case of external irradiation is given by Hummer (1982), 
and for the case of accretion disks and unequal mean opacities 
by Hubeny (1990).

The procedure is as follows. From equation (\ref{reint}) we have
$B=(\kappa_J/\kappa_B) J$, which allows us to express $T$ through
$J$ using the well-known relation $B=(\sigma/\pi) T^4$. To determine $J$, we use
the solution for the second moment of the transfer equation
$K(\tau_H) = H \tau_H + K(0) = 
(\sigma/4\pi)\, T_{\rm eff}^4\, \tau_H + K(0)$,
where $\tau_H$ is the optical depth associated with the flux-mean opacity,
and express the moment $K$ in terms $J$ using the Eddington
factor, $f_K \equiv K/J$. Similarly, we express the surface flux through
the second Eddington factor, $f_H \equiv H(0)/J(0)$,
and we end up with (see also Hubeny 1990)
\begin{equation}
\label{tgen}
T^4 = \frac{3}{4}\, T_{\rm eff}^4\, \frac{\kappa_J}{\kappa_B}
\left[ \frac{1}{3 f_K} \tau_H + \frac{1}{3 f_H} \right] +
\frac{\kappa_J}{\kappa_B}\, W\, T_\ast^4\, .
\end{equation}
Again, this solution is exact within LTE. The usual LTE-gray model
consists in assuming that all the mean opacities are equal to the
Rosseland mean opacity. Moreover, if one adopts the Eddington approximation
($f_K = 1/3; f_H = 1/\sqrt 3$), then one obtains a simple expression
\begin{equation}
T^4 = \frac{3}{4}\, T_{\rm eff}^4 
\left(\tau + 1/\sqrt 3 \right) + W T_\ast^4\, .
\end{equation}
We will consider the most interesting case, namely that of strong irradiation, 
defined by $W T_\ast^4 \gg T_{\rm eff}^4$. In this case, the second
term in brackets is negligible, and we may define a {\em penetration depth} as
the optical depth where the usual thermal part ($\propto T_{\rm eff}^4$)
and the irradiation part ($\propto W T_\ast^4$) are nearly equal, viz.
\begin{equation}
\tau_{\rm pen} \approx W\, \left(\frac{T_\ast}{T_{\rm eff}}\right)^4\, .
\end{equation}
The behavior of the local temperature in the case of the strictly gray model
is very simple -- it is essentially constant, $T= T_0\equiv W^{1/4} T_\ast$ for
$\tau < \tau_{\rm pen}$, and follows the usual distribution
$T \propto \tau^{1/4} T_{\rm eff}$ in deep layers, 
$\tau > \tau_{\rm pen}$.
We stress that while the strictly gray model exhibits an essentially
isothermal structure down to $\tau \approx \tau_{\rm pen}$, such will
not be the case for a nongray model,
as we shall show in detail in the next sections.

\subsection{Surface Layers}
\label{temp_surf}

In the general case, we have to retain the ratio of the
absorption and Planck mean opacities. In the irradiation-dominated
layers ($\tau_H < \tau_{\rm pen}$), the temperature is given by
\begin{equation}
\label{tupp}
T = \gamma\, W^{1/4}\, T_\ast\, ,
\end{equation}
where 
\begin{equation}
\gamma \equiv ({\kappa_J}/{\kappa_B})^{1/4}\, .
\end{equation}
As stated before, $\gamma \approx 1$ in the case of no or weakly irradiated
atmospheres. However, in the case of strong irradiation,  $\gamma$
may differ significantly from unity, and, moreover, may be a strong
function of temperature, and to a lesser extent density. This is easily seen by noticing
that in optically thin regions, the local mean intensity is essentially
equal to twice the irradiation intensity, because the incoming
intensity is equal to the irradiation intensity, and the outgoing 
intensity is roughly
equal to it as well. The reason is that in order to conserve the
total flux when it is much smaller than the partial flux in the inward
or the outward direction, both fluxes should be almost equal,
as are the individual specific and mean intensities. 
More specifically,
\begin{equation}
H = H^{\rm out} - H^{\rm ext} = \int_0^\infty \int_0^1 I_{\nu\mu}\, 
\mu\, d\mu\, d\nu\, -\,
\int_0^\infty \int_0^1 I_{\nu, -\mu}\, \mu\, d\mu\, d\nu\, .
\end{equation}
If $H \ll H^{\rm ext}$, then we must have 
$\int_0^\infty I_{\nu\mu}\, d\nu \approx
\int_0^\infty I_{\nu, -\mu}\, d\nu$ for all angles, and thus 
$J_\nu \approx  2 J_\nu^{\rm ext} \approx 2\,W^{1/4} B_\nu(T_\ast)$.
We may then write the absorption mean opacity as a function of $T$ and $T_\ast$
as
\begin{equation}
\label{abj2}
\kappa_J(T, T_\ast) \approx \frac{\int \kappa_\nu(T) W B_\nu(T_\ast)\, d\nu}
{\int W B_\nu(T_\ast) d\nu} =
\frac{\int \kappa_\nu(T) B_\nu(T_\ast) d\nu}{\int B_\nu(T_\ast)\, d\nu}
\, .
\end{equation}
The dilution factor cancels out, and only the spectral
distribution of the irradiation intensity matters.

In other words, the absorption mean is an average of the opacity 
weighted by the Planck
function corresponding to the effective temperature of the source of
irradiation, while the Planck mean is an average of the opacity 
weighted by the Planck function corresponding to the {\em local} 
temperature. Obviously, they can
be quite different. If the monochromatic opacity differs significantly
in the region where $B_\nu(T)$ and $B_\nu(T_\ast)$ have their local maxima,
the resulting $\kappa_J$ and $\kappa_B$ will differ substantially.
If, moreover, and this is the crucial point, the monochromatic opacity
depends sensitively on temperature, the opacity ratio  $\gamma$
may have a complex, and generally non-monotonic, dependence on
temperature.

The local temperature in the upper layers is given, in view of
equation (\ref{tupp}), by the expression
\begin{equation}
\label{tupp2}
T/T_0 = \gamma(T)\, .
\end{equation}
It is now clear that if $\gamma$ exhibits a strongly non-monotonic behavior
in the vicinity of $T_0$, for instance if it has a pronounced minimum or
maximum there, equation (\ref{tupp2}) may have {\em two or even more
solutions}!

\subsection{Deep Layers}
\label{temp_deep}

The bifurcation behavior is not limited to the surface layers, but
may continue to large optical depths. 
This might seem surprising at first sight because from Equation
(\ref{tgen}) one may expect that once $\tau > 1$, the mean opacities
$\kappa_J$ and $\kappa_B$ become roughly equal, and the
Eddington factor is $f_K \approx 1/3$, so that $T=T_0$ all the way
till $\tau_{\rm pen}$. We recall that the penetration depth may be
quite large; for instance for the case $T_{\rm eff}=75$ K,
$T_\ast= 6000$ K, and
$W=2.2 \times 10^{-3}$ (the case studied in detail in the next
Section), $\tau_{\rm pen} = 9 \times 10^4$.

However, the behavior of the local temperature may be more
complicated. This is linked to another interesting inequality of 
mean opacities that is usually taken for granted, namely the flux 
mean opacity $\chi_H$ and the Rosseland mean opacity $\chi_{\rm ross}$.
The Rosseland mean opacity is in fact defined in such a way that it is
equal to the flux mean opacity in the diffusion approximation.
Indeed, in this approximation,
\[
H_\nu \approx \frac{1}{3}\frac{dB_\nu}{d\tau_\nu} = 
\frac{1}{3}\frac{1}{\chi_\nu}\frac{dB_\nu}{dm}= 
\frac{1}{3}\frac{1}{\chi_\nu}\frac{dB_\nu}{dT}\frac{dT}{dm}\, ,
\]
and therefore
\[
\chi_H \approx \frac{\int_0^\infty\frac{dB_\nu}{dT}d\nu}
{\int_0^\infty\frac{1}{\chi_\nu}\frac{dB_\nu}{dT}d\nu} \equiv 
\chi_{\rm ross}\, ,
\]
where the latter equality represents the definition of the
Rosseland mean opacity.

In the case of strong irradiation, an interesting, and fundamentally
different, situation appears. Since the net flux is very small,
the total flux in the $\mu > 0$ (outgoing) hemisphere is roughly
equal to that in the $\mu <0$ (incoming) hemisphere (see above).
Because the monochromatic opacity varies strongly with frequency,
there are frequency regions where the net monochromatic flux is
positive, and regions where it is negative. The flux-mean opacity
close to the surface may attain large values, either positive
or negative, depending upon whether $\chi_\nu$ 
weighs the positive or negative net flux regions more.

Going to deeper layers, the net monochromatic flux decreases because the
radiation field becomes more isotropic for all frequencies, so that
the flux mean opacity decreases. Consequently, the corresponding
flux-mean optical depth increments 
$\Delta \tau_H \approx \chi_H \Delta m$
become very small, and $\tau_H$ will exhibit a plateau where it
remains essentially constant with $m$. Finally, in the regions
where all the influence of the external irradiation dies out,
the usual diffusion approximation sets in, and the flux-mean optical
depth becomes essentially equal to the Rosseland optical depth.

This means that for strong irradiation, the temperature in the
deep layers should exhibit a plateau with
\begin{equation}
T_{\rm plateau} \approx \frac{3}{4}\, T_{\rm eff}^4\, \bar\tau_H\, ,
\end{equation}
where $\bar\tau_H$ is the plateau value of the flux-mean optical depth.
The value of $\bar\tau_H$ and thus $T_{\rm plateau}$ depends upon
the exact form of the monochromatic opacity; a rough estimate
is provided by setting $\bar\tau_H \approx \tau_{\rm pen}$, in which
case $T_{\rm plateau} \approx (7/4)^{1/4}W^{1/4} T_\ast
\approx 1.15\, T_0$.


\section{Example of an Extrasolar Giant Planet Irradiated by a 
Solar-type Star}

\subsection{Opacities}

Using partition functions, LTE level densities, stimulated emission
corrections, and broadening algorithms, we generated opacity
tables from numerous available line lists. 
For gaseous H$_2$O, Partridge and Schwenke (1997) 
have calculated the strengths of more than $3\times 10^8$ lines. We used
a subset of their $4\times 10^7$ strongest lines.
For various other molecular species ({\it e.g.}, NH$_3$, PH$_3$, H$_2$S and CO),
we used the HITRAN (Rothman \etal\ 1992,1998) and GEISA (Husson \etal\ 1994)
databases, augmented with additional lines from theoretical
calculations and measurements (Tyuterev \etal\ 1994; Goorvitch 1994; 
Tipping 1990; Wattson and Rothman 1992). For methane, shortward of $\sim$1.0 \mic we used the 
Karkoschka (1994) opacities and between $\sim$1.0 \mic and 1.58 \mic we used the Strong \etal (1993)
opacities. Longward of 1.58 \mic, we used the Dijon methane database (Borysow et al. 2003)
that includes the hot bands. Opacity due to Collision--Induced 
Absorption (CIA) by H$_2$ and helium is taken from Borysow and
Frommhold (1990), Zheng and Borysow (1995), and Borysow, J\o{rgensen}, and Zheng (1997), as updated
in Borysow (2002). 

FeH and CrH opacities were taken from Dulick et al. (2003) and Burrows et al. (2002), respectively.
The TiO line lists for its nine major electronic systems were taken from the Schwenke ab initio
calculations,
as modified by Allard, Hauschildt, and Schwenke (2000). These data include 
lines due to isotopically substituted molecules $^{46}$Ti$^{16}$O, $^{47}$Ti$^{16}$O,
$^{49}$Ti$^{16}$O and $^{50}$Ti$^{16}$O relative to the most abundant
isotopic form $^{48}$Ti$^{16}$O.
For VO, we used the line list provided by Plez (1999). Because
$^{51}$V is by far vanadium's most abundant isotope, the lines of isotopically
substituted molecules are not necessary.
The line lists and strengths for the neutral alkali elements (Li, Na, K, Rb and Cs) were obtained
from the Vienna Atomic Line Data Base (Piskunov \etal 1995).
The general line shape theory of Burrows, Marley, and Sharp (2000)
was used for Na and K, while those of Nefedov \etal (1999) and 
Dimitrijevi\'{c} and Peach (1990) were used for the other alkali metals.

In practice, we pre-calculate a large opacity table in $T$/$\rho$/frequency space
in which we interpolate during the iteration of the atmosphere/spectral model
to convergence. The table contains 30,000 frequency points from 0.3 to 300 \mic, 
uniformly spaced in steps of 1 cm$^{-1}$. Generally, 5000 geometrically-spaced 
frequency points are used in the radiative transfer model. In other words,
our approach belongs to the category of ``opacity sampling'' methods for
treating line blanketing.

We consider two different opacity tables; one which 
includes opacity due to TiO and VO, and the other one without these
molecules. In the case without TiO/VO, these species were removed both
from the opacity table as well as from the state equation, so these species
are considered as completely absent everywhere in the atmosphere.
These two different opacity tables suggest themselves because for
an irradiated atmosphere we obtain high temperatures in the low pressure
outer atmosphere. Importantly, temperature inversions are very possible.
However, thermochemical equilibrium calculations with rainout 
(Burrows \& Sharp 1999;
Lodders  \& Fegley 2002) suggest that TiO and VO would exist at 
such high-$T$/low-$P$ points.  Since the presence of TiO and VO is 
quite natural at high-$T$/high-$P$ points, one confronts a situation 
in which there are two regions rich in TiO/VO, separated
by a middle region that is TiO/VO free. In a gravitational field with a monotonic
pressure profile, this gap could act like a cold trap in which the TiO/VO that is
transported by molecular or eddy diffusion from the upper low-P region into the
intermediate cooler region, would condense and settle out, thereby depleting the
upper low-$P$ TiO/VO-rich region. This would eventually leave no TiO/VO at altitude to provide
the significant absorption that could lead to a bifurcation. However, the cold-trap effect can not
be anticipated in abundance tables that are functions of temperature and pressure alone.
These tables are not cognizant of gravity, nor are they aware of the global 
$T$/$P$
profile. Whether this cold-trap effect in fact happens is not yet clear. Furthermore,
the same cold-trap effect might obtain for other EGP species (e.g. Fe). Hence,
we use the presence or absence of TiO/VO at high-$T$/low-$P$ points and the associated ambiguity
as a means to explore the real mathematical bifurcation effect we have identified
in this paper and leave to a future work the study of the true viability of
cold traps in irradiated EGP atmospheres.

\subsection{Results}

We consider an intrinsically cold giant planet, with $T_{\rm eff} = 75$ K,
$\log g = 3$,
irradiated by a G0V star with a spectral energy distribution taken 
for simplicity to be the corresponding
Kurucz (1994) model atmosphere. We consider the class V, ``roaster''
situation, as defined by SBH.

\begin{figure}
\plotone{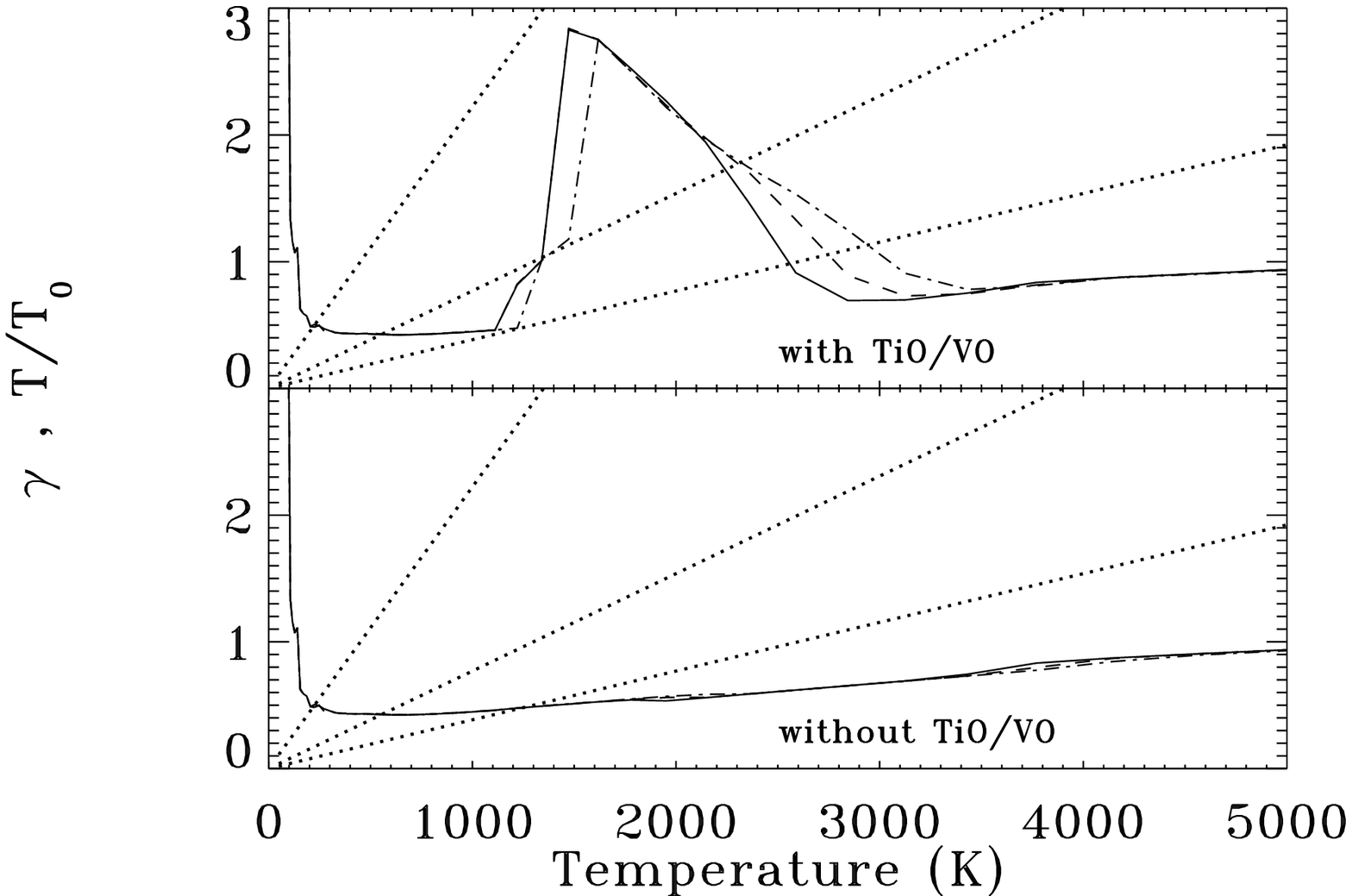}
\caption[]{Opacity ratio index $\gamma=(\kappa_J/\kappa_B)^{1/4}$ plotted 
as a function of temperature. The upper panel shows the results for the opacities
including TiO/VO, while the lower panel shows those without TiO/VO.
The absorption mean opacity was evaluated
using equation (\ref{abj2}) with the stellar effective temperature
$T_\ast = 6000$~K. The full line corresponds to a density $\rho$ of
$10^{-12}$ g cm$^{-3}$;
the dashed line to $\rho=10^{-11}$ g cm$^{-3}$; and the dot-dashed 
line to  $\rho=10^{-10}$ g cm$^{-3}$.
The dotted lines represent the quantity $T/T_0$, with $T_0$ corresponding
to three different distances from the star. From top to bottom,
$T_0=450$ K, (0.65 AU), $T_0=1250$ K (0.08 AU), and
$T_0=2600$ K,  (0.0225 AU). The intersections of the individual dotted lines
with the $\gamma(T)$ curve represent possible solutions for the
temperature in the upper layers of an irradiated atmosphere.
In the case of the opacity table with TiO/VO, 
there are unique solutions for low and for high 
irradiation (corresponding to $T\approx 250$ K and
$T \approx 2600 - 3000 K$), while in the intermediate
cases there are indeed multiple solutions. 
The opacity table without TiO/VO does not exhibit a peak around 1500 K,
so one obtains unique solutions for the temperature for all irradiations.
}
\label{fig1}
\end{figure}

The opacity ratio index $\gamma$ is plotted in Figure~\ref{fig1} as a function
of temperature, for three densities characteristic to the outer
layers of an atmosphere -- $\rho = 10^{-12}$, $10^{-11}$, and $10^{-10}$
g cm$^{-3}$. The upper panel shows the results for the opacities
including TiO/VO, while the lower panel shows those without TiO/VO.
It is clearly seen that in the former case the ratio $\kappa_J/\kappa_B$,
and thus $\gamma$, exhibits a sharp peak around 1500 K; the dependence
on density is not strong. The ratios $T/T_0$ for $T_0=450$ K (weak 
irradiation), 1250 K (medium strong irradiation), and 2600 K 
(very strong irradiation) are overplotted with dotted lines. The latter
roughly corresponds to the case of the recently discovered transiting
planet OGLE-TR-56. According to equation (\ref{tupp2}), the temperature
at the surface layers is found at the intersection of the curves  of
$\gamma(T)$ and $T/T_0$; thus, we see that for low and for high irradiation
there are unique solutions (corresponding to $T\approx 250$ K and
$T \approx 2600 - 3000 K$, but that in the intermediate
cases there are indeed multiple solutions. 

The opacity table without TiO/VO does not exhibit a peak around 1500 K,
and as it is clearly seen from the lower panel of Fig.~\ref{fig1},
one obtains unique solutions for the temperature for all reasonable
irradiations. Mathematically, the case of truly 
extreme irradiation  (e.g. with $T_0$ around 4700 K) might again lead 
to multiple solutions\footnote{We are indebted to the referee for pointing 
out this possibility to us.}
because the line $T/T_0$ now almost coincides with $\gamma(T)$. However,
we do not consider such a case here because then a host of new phenomena
which we do not address in our present model, such as departures from LTE, 
dynamical effects, etc., would likely become crucial.

\begin{figure}
\plotone{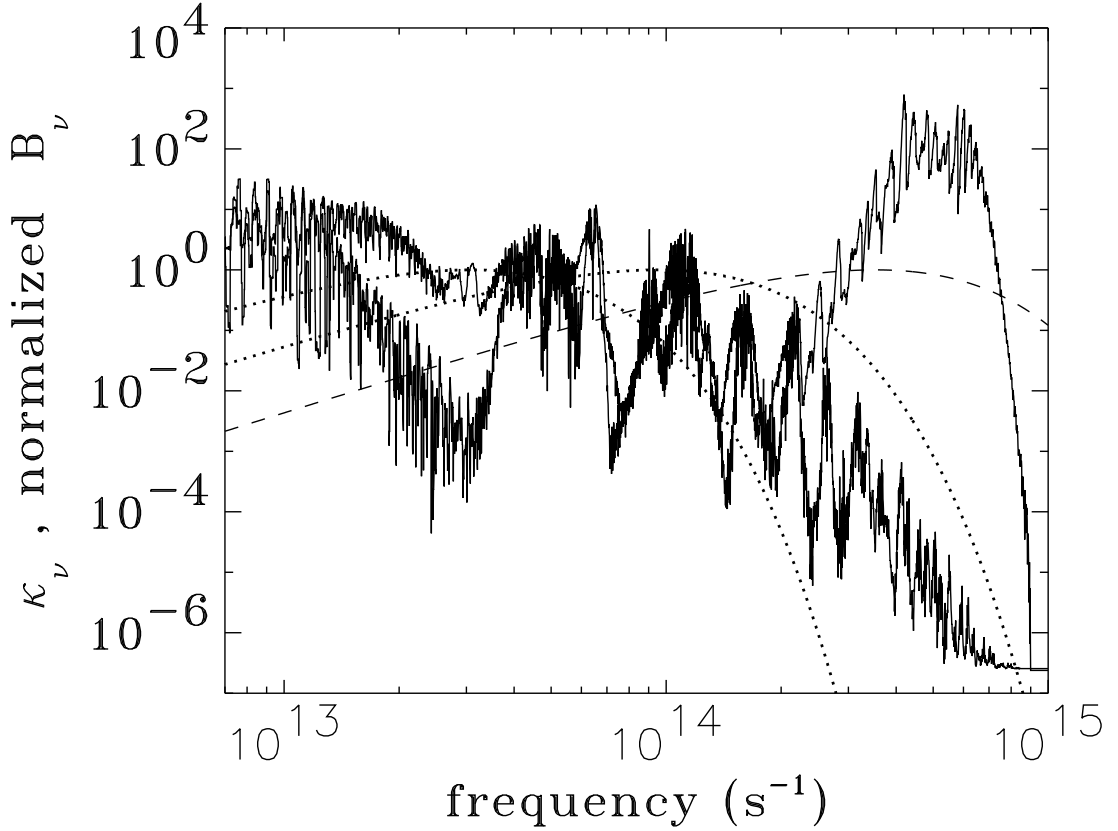}
\caption[]{Monochromatic opacity (in cm$^{2}$ g$^{-1}$) as a function 
of frequency for density $\rho = 10^{-11}$ g cm$^{-3}$ and for 
two temperatures, $T = 1600$ K (solid line, with a maximum around
$\nu \approx 3 - 6 \times 10^{14}$ Hz), and $T=520$ K (bold solid line);
together with normalized Planck functions for $T=T_\ast$ (dashed line),
and for the two nominal temperatures (1600 and 520 K -- dotted lines).
A large monochromatic opacity for $T=1600$ K in the optical region where 
$B_\nu(T=T_\ast)$ has a maximum explains why $\kappa_J > \kappa_B$ for
this temperature.
}
\label{fig0}
\end{figure}

To explain the behavior of the $\kappa_J$ over $\kappa_B$ ratio,
we plot in Figure~\ref{fig0} two examples of the monochromatic opacity, 
for $T = 1600$ K (close to the maximum of $\gamma$),
and $T = 520$ K (the region near the minimum of $\gamma$). In both cases
$\rho = 10^{-11}$ g cm$^{-3}$. We also plot normalized
Planck functions for the two nominal temperatures, 1600 and 520 K
(dotted lines), and for $T=T_\ast=6000$ K (dashed line). 
The latter weighs most heavily the optical region
around $\nu \approx 4 - 5 \times 10^{14}$ Hz, 
while the local Planck functions put the weight at much
lower frequencies (around $10^{14}$ and $3 \times 10^{13}$ Hz,
respectively). In the case of the higher temperature, $T=1600 $ K,
the monochromatic opacity exhibits a maximum just in the region
where the Planck function corresponding to $T_\ast$ is largest 
(because of TiO and VO); therefore, $\kappa_J > \kappa_B$.
For the lower temperature, $T=520$ K, the opacity in the optical range
is lower by orders of magnitude, so $\kappa_J$ is now significantly
lower. Since the Planck function for $T=520$ K emphasizes frequencies
below $7 \times 10^{13}$ Hz, where the monochromatic opacity is very large,
we obtain $\kappa_J < \kappa_B$.

There are two other features seen in Figure~\ref{fig0} that are worth noticing.
First, there is a large increase of $\gamma$ for temperatures around
and below 100 K. This is a consequence of low opacity at low frequencies. 
However, the opacity sources are quite uncertain in this regime. For studies
of extremely cool objects the opacity table should be checked and
possibly upgraded, but studying these objects is not our objective here.
Second, in the case of no TiO/VO opacity (lower panel), it appears that
$\gamma$ nearly coincides with  $T/T_\ast$. Taken at the face value,
this would imply that the ratio 
${\int \kappa_\nu B_\nu(T_{\star}) d\nu}/{\int \kappa_\nu B_\nu(T) d\nu}$ 
is very close to unity.\footnote{We are indebted to the referee for drawing our
attention to this point.} However, we do not see any compelling
reason why this should be generally valid, so we conclude that such a behavior
is just a coincidence.

\begin{figure}
\plotone{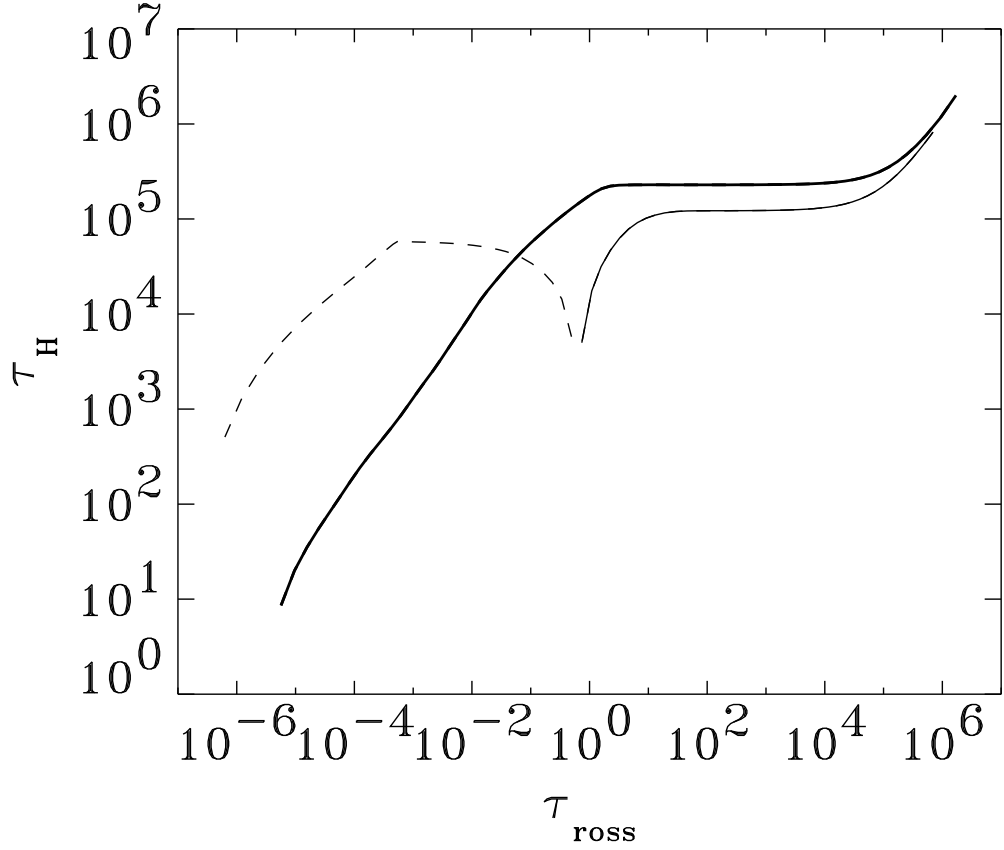}
\caption[]{Flux-mean optical depth as a function of Rosseland mean 
optical depth
for the two solutions of the
intermediate-irradiation (distance 0.08 AU) model, computed
with the TiO/VO opacity. The heavy line represents the low-surface-$T$ branch;
the thin line the high-surface-$T$ branch. The solid line represents
the positive values of $\tau_H$, while the dashed line the negative 
values.
}
\label{figtauh1}
\end{figure}

In Figure~\ref{figtauh1} we demonstrate the behavior of the
flux-mean opacity discussed in \S\ref{temp_deep}. The high-surface-$T$ 
branch exhibits a negative flux mean optical depth at the surface layers
because the opacity is high in the optical region where the next flux 
is negative. Consequently, the flux-mean opacity is negative there,
and thus the flux-mean opacity (and the corresponding optical depth)
on the plateau is lower than in the case of the low-surface-$T$ branch.

We have also computed exact LTE models as described in SBH. 
We have used the stellar atmosphere code COOLTLUSTY, which is a variant
of the universal program TLUSTY (Hubeny 1988; Hubeny \& Lanz 1995).
We stress that the emergent flux is not computed by a separate program;
instead it is computed already by COOLTLUSTY. This also means that we 
do not use different opacity tables for computing the atmospheric structure 
and for computing the spectra. We have also tested the sensitivity of the
computed model on different opacity samplings; we have resampled our
original frequency grid of 5000 points to lower numbers of frequencies, 
and even for 300 points we found very little differences in the temperature
structure (of the order of a few K). In some cases it was found advantageous
to converge first a model with 300 frequencies, and using this as an
input to reconverge a model with the full grid of 5000 frequencies in the
next step.

\begin{figure}
\plotone{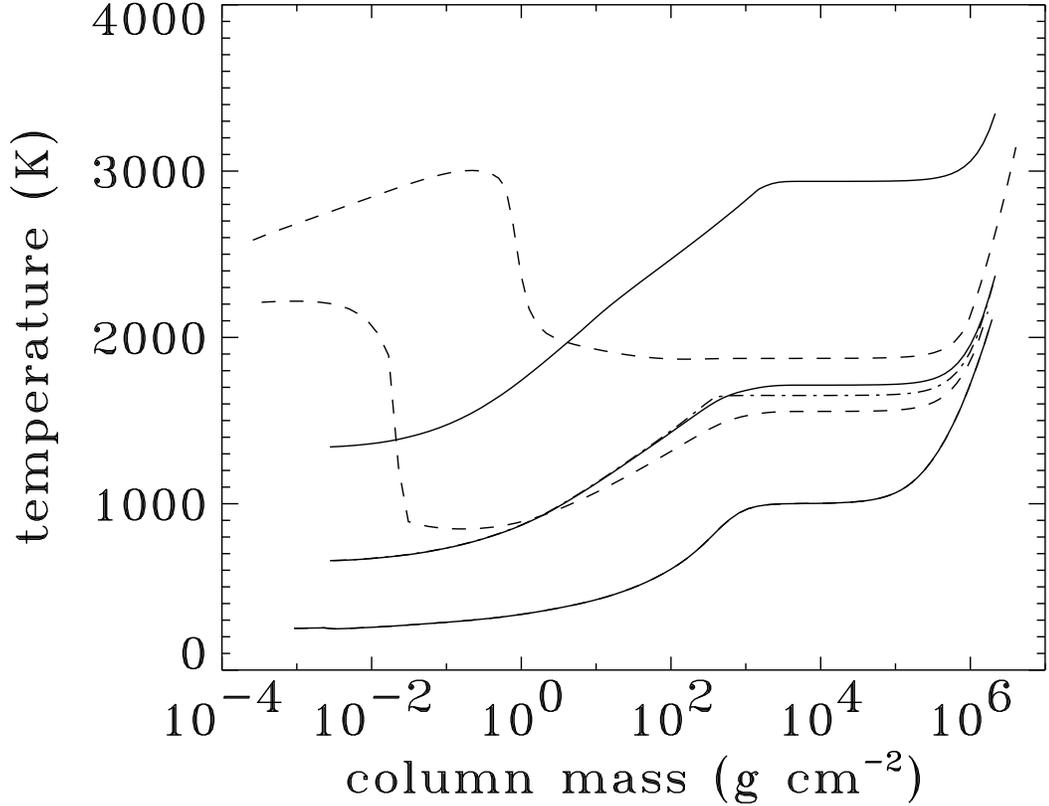}
\caption[]{Temperature as a function of column mass.
Solid lines represent models computed for the opacity table without
TiO/VO, while the dashed and dot-dashed lines represent models 
with TiO/VO. The curves represent, from top to bottom,
three distances from the central star.
The top lines corresponds to 0.0225 AU, the middle lines to 0.08 AU,
and the bottom lines to 0.65 AU. Numerical values of the dilution
factor are $W = 2.8 \times 10^{-2}$, $ 2.2 \times 10^{-3}$, and 
$3.5 \times 10^{-5}$.
The models for the lowest irradiation ($W =3.5 \times 10^{-5}$)
computed with and without TiO/VO are not distinguishable on the plot.
The middle dashed and the dot-dashed lines 
represent the two solutions for the same distance, and thus
illustrate the bifurcation discussed in the text. 
}
\label{figtemp1}
\end{figure}

The temperature
structure is displayed in Figure~\ref{figtemp1}, which demonstrates that the
simplified description put forward above faithfully reflects the true behavior
of temperature. 
Full lines represent models computed for the opacity table without
TiO/VO, while the dashed and dot-dashed lines represent models 
with TiO/VO. The models without TiO/VO always converged to a unique
solution, moreover with temperature monotonically decreasing
outward, as we may have expected from the above discussion.

The behavior of the models with TiO/VO is more interesting.
The high-irradiation model, the top dashed line, indeed converged to a unique
solution. The surface temperature is 2600 K, in excellent agreement
with the value expected from Figure~\ref{fig1} (the density at the uppermost
point is about $3 \times 10^{-12}$ g cm$^{-3}$). Even the initial increase
of temperature in the upper layers when going inward is explained by 
Figure~\ref{fig1}, since for 
progressively higher densities the solution of $T/T_0=\gamma(T)$ occurs
at higher temperatures. The low-surface-$T$ solution is very close to
the no-TiO/VO model, which once again demonstrates that the
bifurcation is caused by the TiO/VO opacity.
We have also examined depth-dependent concentrations of TiO and VO. Indeed,
the mole fraction of TiO drops precipitously from $\approx 3\times 10^{-7}$
to essentially zero at $m \approx 1.6 \times 10^{-2}$ g cm$^{-3}$ (that is at
pressure $\approx 1.6 \times 10^{-5}$ bars and temperature $T \approx 1500$ K)
for the high-surface-$T$ model; while is is not present in the low-surface-$T$ 
model. Behavior of the VO concentration is analogous.

The low-irradiation model also converged to
a unique solution; moreover, its surface temperature of about 250 K is 
in excellent agreement with the value one may expect from 
Figure~\ref{fig1}. Again, both models, without and with TiO/VO agree
very well.

The two middle curves, the dashed and dot-dashed curves, correspond
to the same irradiation, and, thus, represent the bifurcation predicted
by our analysis. The model with high surface temperature was converged
starting from the extremely irradiated model (through several intermediate
steps), while the model with the low surface temperature was converged
starting from scratch, i.e. from a usual LTE-gray model in which all mean
opacities are created equal. The high surface temperature at about 2200 K 
essentially coincides with the high-temperature intersection of about 2200 K
on Figure~\ref{fig1}. The low-surface-$T$ solution exhibits a somewhat
higher temperature than would follow from the low-$T$ intersection in
Figure~\ref{fig1}  -- 650 versus 550 K, but in view of all the approximations
involved such an agreement is still very good. We did not numerically
find the third possible solution with a surface $T$ of about 1300 - 1400 K.

We stress that the numerical solutions were not contrived in any way to
lead to the predicted bifurcation. In fact, we first discovered the
effect purely numerically -- we obtained two different, yet perfectly well
converged solutions, depending on the input model. We have continued
the convergence of both solution until the maximum relative change of
temperature and density decreased below $10^{-5}$ in all depth points.

Several features of the overall accuracy of the model are worth mentioning.
In fact, the low value of maximum relative change of temperature 
and density is not a satisfactory criterion of the convergence. One should also
check the accuracy of the computed flux gradient and the value of the net
flux. That is, for the former case  we have to examine the quantity 
$e_{\rm grad} = (\kappa_J J - \kappa_B B)/\kappa_B B$, and for the
latter case the quantity 
$e_{\rm flux} = (H - H_0)/H_0)$, where $H_0 = (\sigma/4\pi) T_{\rm eff}^4$
is the net flux. For the models presented in Figure~\ref{figtemp1},
$e_{\rm grad}$ is very small, typically $10^{-10}$ or less. The conservation
of the total net flux is much harder to achieve; the bifurcated models
have typically $e_{\rm flux} \approx {\rm few} \times 10^{-4}$ near the
surface, while $e_{\rm flux}$ reaches a few $\times 10^{-2}$ for column mass 
around $ m \approx 10^2$ g cm$^{-2}$ (that is, a few per cent), and
drops to less than $10^{-5}$ at $m \approx 10^4$  g cm$^{-2}$ and deeper. 
The error
in the net total flux of several per cent might seem large, but one has
to bear in mind that the net flux is a subtraction of two large quantities,
$H^{\rm out} - H^{\rm in}$, and the actual numerical error
$(H^{\rm out} - H^{\rm in} - H_0)/H^{\rm out}$ is of the order of $10^{-5}$
or less. A dramatic demonstration of this fact is that a model, which
is almost converged (maximum relative change in $T$ and $\rho$ was less
than $10^{-3}$), exhibits a seemingly ridiculous value of
$e_{\rm flux} \approx  10^5$ at upper layers $m < 10^2$ g cm$^{-2}$ 
(i.e. an error in the total net flux of some $10^7$\% !), yet the actual
temperature difference between this and the fully converged model is
it most about 2-3 K, and the predicted flux in both cases
is completely indistinguishable on a plot! This again shows that in the
case of strong irradiation it is the value of 
$(H^{\rm out} - H^{\rm in} - H_0)/H^{\rm out}$ (or $e_{\rm grad}$)
which is critical for practical convergence, not a much more stringent 
criterion of the error in the total net flux, $e_{\rm flux}$.

\begin{figure}
\plotone{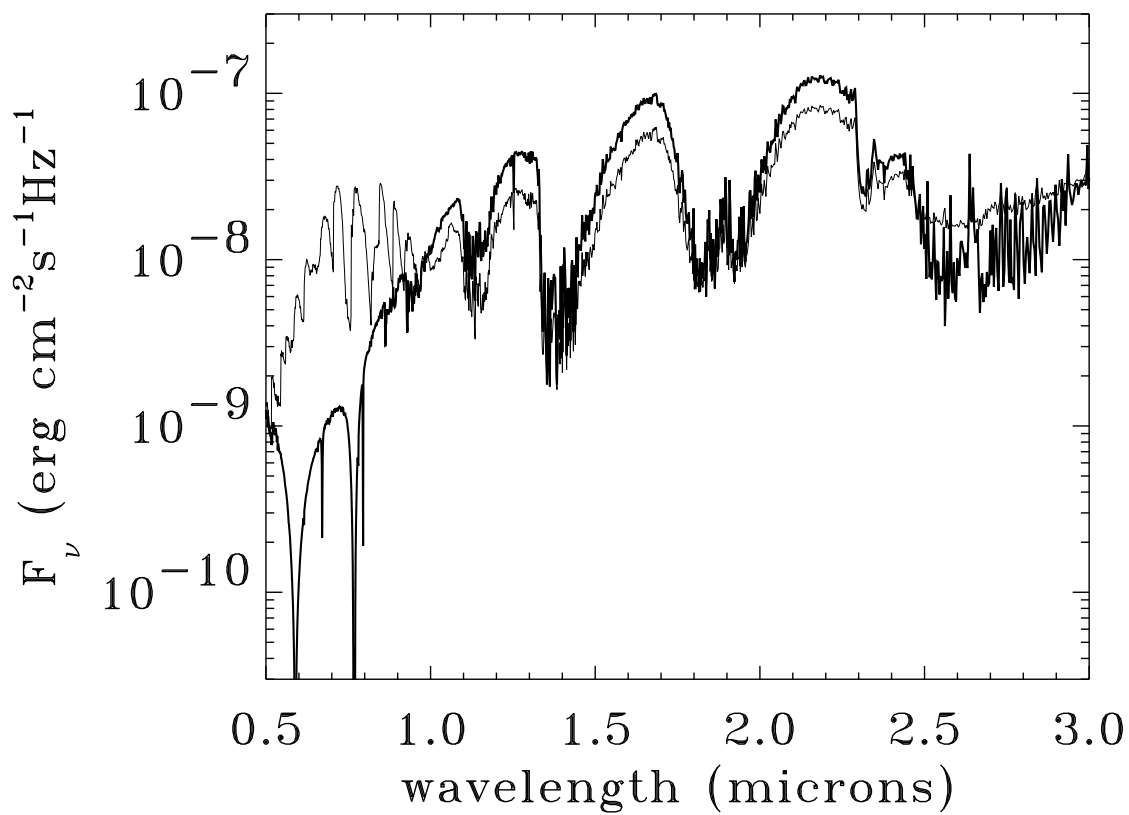}
\caption[]{Emergent flux of a function of wavelength for
the two solutions of the
intermediate-irradiation (distance 0.08 AU) model, computed
with TiO/VO. Heavy line represents the low-surface-$T$ branch;
the thin line the high-surface-$T$ branch. The model flux computed
without TiO/VO is very close to the low-surface-$T$ branch,
and for clarity is not displayed.
}
\label{figflux1}
\end{figure}

The difference in the temperature structure is of course reflected
in the emergent spectrum. In Figure~\ref{figflux1} we show the
emergent flux corresponding to the two solutions of the
intermediate-irradiation (distance 0.08 AU) model, computed
with TiO/VO. The models differ dramatically in the optical and in near-IR
region because of the dramatically different surface temperature
and the corresponding increase of the TiO/VO opacity.
Since the low-surface-$T$ branch exhibits a higher
temperature at the plateau (see Fig.~\ref{figtemp1}), the flux
at the low-opacity regions is higher.

\begin{figure}
\plotone{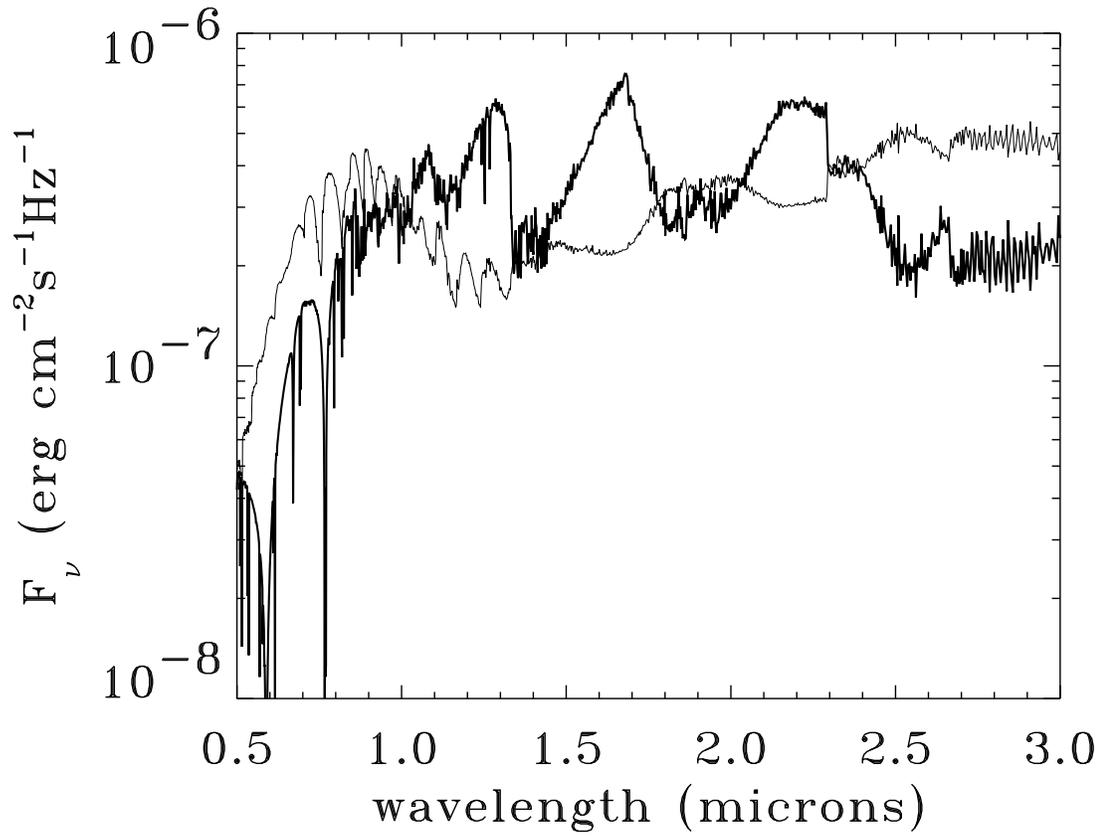}
\caption[]{Emergent flux of a function of wavelength for
the high-irradiation (distance 0.0225 AU) model, computed
without TiO/VO (thick line) and with TiO/VO (thin line).
}
\label{figflux2}
\end{figure}

Finally, Fig.~\ref{figflux2} displays the emergent flux for
the two models for the highest irradiation, namely for the distance 
0.0225 AU, which corresponds to OGLE-TR-56. The thick line is a model 
without TiO/VO, and the thin line with TiO/VO. There is no 
bifurcation here since both opacity tables led to unique solutions,
however quite different ones. The corresponding spectra are thus
significantly different as well.

\section{Discussion and Conclusions}

We have demonstrated that under certain circumstances the differences
between the absorption mean and the Planck mean opacities can lead to
multiple solutions for the atmospheric structure. This result is
quite robust, and in fact represents a so far overlooked general result
of elementary stellar atmospheres theory. Since we do not expect
that the absorption and Planck mean opacities will differ significantly
in the usual case of radiative equilibrium of non-irradiated atmospheres,
the most interesting situations where the effect may play role are
strongly irradiated stars and planets, and also, possibly, structures
where there is a significant deposition of mechanical energy, such as
stellar chromospheres and accretion disks. We are not concerned
with these objects here, but would like to stimulate further study of 
possible bifurcations for these situations.
We note that Nolan \& Lunine (1988) also found a bimodal behavior
of the atmosphere of Triton which is linked to external
irradiation.

From the physical point of view we may interpret the absorption
mean opacity as the global absorption efficiency, 
and the Planck mean opacity as the global emission efficiency,
of the medium. The integrated mean intensity $J$ acts as a total
absorption pool, while the integrated Planck function $B$ as a total
thermal pool. The radiative equilibrium, that stipulates that the
total radiation energy absorbed on the spot is equal to the total
energy emitted on the same spot, 
therefore acts as a thermostat: The radiative equilibrium sets the
local temperature in such a way that $\kappa_J J = \kappa_B B$.
Here $J$ is determined by the radiative transfer, and the local 
temperature follows as 
$B \equiv (\sigma/\pi) T^4 = (\kappa_J/\kappa_B)\, J$. In the
case of strong irradiation, the weighting that determines $\kappa_J$
is dominated by the incoming radiation. If the monochromatic opacity
in the region around the peak of the external irradiation is very
sensitive to temperature, for instance if it is very high for
high temperatures and very low for low temperatures, the
``radiative-equilibrium thermostat'' may find two solutions -- either
a high $T$ together with high $\kappa_J$, or a low $T$ with low
$\kappa_J$; in both cases the radiative equilibrium 
$\kappa_J J = \kappa_B B$ is satisfied.

However, the applicability of 
this effect to real objects depends critically
on a degree of reliability of adopted opacities. For high-temperature
conditions where the gaseous opacities dominate, the situation is
relatively well under control, thanks to recent enormous progress
in computing atomic data (Opacity Project, Iron Project, OPAL, and others;
for a recent review see, e.g., Nahar 2003).

In the case of irradiated giant planets or other substellar-mass objects
like L and T dwarfs, the opacity for low density
($\rho \approx 10^{-12} - 10^{-9}$ g cm$^{-3}$)
and high temperature ($T > 2000$ K) is still relatively uncertain.
At the present stage, we cannot be sure that the bifurcation effect
really occurs, but we have shown that this is certainly a very real
possibility which should be taken into account when constructing
atmospheric models.

\acknowledgments

The authors are pleased to thank Bill Hubbard, Thierry Lanz,
Jonathan Lunine, Jim Liebert, Christopher Sharp, and Drew Milsom
for fruitful conversations and help during the
course of this work, as well as
NASA for its financial support via grants NAG5-10760
and NAG5-10629. We also thank anonymous referee for a number of
comments and suggestions.

\end{document}